\documentclass[twocolumn,showpacs,preprintnumbers,amsmath,amssymb]{revtex4}
\usepackage{dcolumn}
\usepackage{bm}
\usepackage{graphicx}

\def\br{{\bf r}}

\def\bk{{\bf k}}
\def\bkp{{\bf k}'}
\def\kvac{\mid \widetilde{0}\rangle}
\def\bvac{\langle \widetilde{0}\mid}
\def\wo{\omega_{\text{osc}}}
\newcommand{\ket}[1]{\left| #1 \right\rangle }

\def\wk{\omega_k}
\def\wj{\omega_j}

\def\wc{\omega_{\text{cut}}}

\begin{document}

\title{Vacuum Casimir energy densities and field divergences at boundaries}

\author{Nicola Bartolo\mbox{$^{1,2}$}, Salvatore Butera\mbox{$^{3}$}, Margherita Lattuca\mbox{$^{4}$}, Roberto Passante\mbox{$^{4}$}, Lucia Rizzuto\mbox{$^{4}$}, and Salvatore Spagnolo\mbox{$^{4}$}}
\affiliation{\mbox{$^{1}$}INO-CNR BEC Center and Dipartimento di Fisica, Universit\`{a} di Trento, I-38123 Povo, Italy
\\
\mbox{$^{2}$}Laboratoire Charles Coulomb (L2C), UMR 5221 CNRS-Univ. Montpellier 2, Montpellier, France
\\
\mbox{$^{3}$}SUPA, Institute of Photonics and Quantum Sciences, Heriot-Watt University, Edinburgh, UK
\\
\mbox{$^{4}$}Dipartimento di Fisica e Chimica, Universit\`{a} degli Studi di Palermo and CNISM, Via Archirafi 36, I-90123 Palermo, Italy}

\email{roberto.passante@unipa.it}


\begin{abstract}
We consider and review the emergence of singular field fluctuations or energy densities at sharp boundaries or point-like field sources in the vacuum. The presence of singular energy densities of a field may be relevant from a conceptual point of view, because they contribute to the self-energy of the system. They could also generate significant gravitational effects. We first consider the case of the interface between a metallic boundary and the vacuum, and obtain the structure of the singular electric and magnetic energy densities at the interface through an appropriate limit from a dielectric to an ideal conductor.
Then, we consider the case of a nondispersive and nondissipative point-like source of the electromagnetic field, described by its polarizability, and show that also in this case the electric and magnetic energy densities show a singular structure at the source position. We discuss how, in both cases, these singularities give an essential contribution to the electromagnetic self-energy of the system; moreover, they solve an apparent inconsistency between the space integral of the field energy density and the average value of the field Hamiltonian. The singular behavior we find is softened, or even eliminated, for boundaries  fluctuating in space and for extended field sources. We discuss in detail the case in which a reflecting boundary is not fixed in space but is allowed to move around an equilibrium position, under the effect of quantum fluctuations of its position. Specifically, we consider the simple case of a one-dimensional massless scalar field in a cavity with one fixed and one mobile wall described quantum-mechanically. We investigate how the possible motion of the wall changes the vacuum fluctuations and the energy density of the field, compared with the fixed-wall case. Also, we explicitly show how the fluctuating motion of the wall smears out the singular behaviour of the field energy density at the boundary.
\end{abstract}

\maketitle

\section{\label{sec:1}Introduction}

Casimir forces, which are electromagnetic interactions between neutral macroscopic metallic or dielectric objects, arise from the change of the zero-point
energy of the field consequent to the change of a boundary of the system considered \cite{CP48,Milonni94}. Such a change modifies the boundary conditions on the field operators and thus the energy associated with the field vacuum fluctuations.

The zero-point (vacuum) field energy density is related to the existence of vacuum field fluctuations, and it may become singular (divergent) at sharp boundaries with metallic bodies \cite{Milton04,CPMK07,MNS11,BP12,BK02}.
The singularity of the energy density gives an essential contribution to the electromagnetic self-energy of a body, and it should also act as a source of the gravitational field \cite{Milton11,MNS11}. The problem about whether fluctuating fields gravitate at all has been questioned in the literature \cite{FMPRSW07}, and it is an important issue concerning with the reality of quantum vacuum fields.   An analogous singular behaviour of the electric and magnetic energy density occurs also in the case of a point-like field source in the vacuum \cite{PRS13}.
These divergences of vacuum energy densities and field fluctuations, also, strongly depend on the (ultraviolet) cutoff on the field modes \cite{EFM12}, and may be model-dependent. They must be handled carefully in order to ensure consistency between local and global Casimir energies, for example in the cases of a flat metal-vacuum interface \cite{BP12,Milton04}  and a spherical or cylindrical shell \cite{MCPW06,CPMK07}. It has been shown that when the boundary has position fluctuations, the singular energy density is removed \cite{FS98}; however, also in this case a strong change of the energy density in the proximity of the boundary or of the field source is present.
Therefore, relevant gravitational effects could be also present; however, as mentioned, it is still debated in the literature whether vacuum field fluctuations can couple to gravity.
 Divergences can be also softened or eliminated using a power-laws  potential to represent the wall \cite{Milton11a}.

In this paper we consider and review the problem of the singular behavior of vacuum electric and magnetic energy densities in three different cases: a perfectly conducting metallic wall, a point-like source of the electromagnetic field, and a 1D system of a massless scalar field confined in a cavity with a fixed boundary and a wall which is free to move and thus has quantum fluctuations of its position. We obtain the field energy densities in the three cases and discuss their singular behaviour at the boundary. Also, we point out the importance of the singular terms to obtain correctly the self-energy of the object considered, as well as consistency between the field self-energy (average value of the field Hamiltonian) and the space integral of the field energy density.

This paper is organized as follows. In Section \ref{sec:2}, using an appropriate limiting procedure from a dielectric to an ideal conductor, we show the presence of singularities of the electric and magnetic energy density associated to vacuum fluctuations in the proximity of the wall, discussing their mathematical structure. In Section \ref{sec:3}, using an analogous procedure, we obtain the structure of the energy density singularity at the position of a dielectric point-like source of the quantum electromagnetic field in its dressed ground state. We show the importance of these singular terms to obtain correctly the electromagnetic self-energy of the source from the space integral of the electromagnetic energy densities. In Section \ref{sec:4} we consider a massless scalar field in a one-dimensional cavity with a fixed and a mobile wall. We discuss how the position fluctuations of the mobile wall change the field energy density inside the cavity compared to the static walls case, and how in this case surface divergences are softened and smeared out by the quantum fluctuating motion of the wall. Finally, Section \ref{sec:5} is devoted to our concluding remarks.

\section{\label{sec:2}Surface divergence of the field energy density near a perfectly conducting metallic wall}

In this Section we consider the electromagnetic vacuum fluctuations in the empty space near a perfectly conducting metallic half-space. Using an appropriate limiting procedure from a dielectric to the ideal conductor limit, we show the presence of singularities of the vacuum electric and magnetic energy densities at the vacuum-metal interface.

We assume the presence of a wall-vacuum boundary at $z=0$ with the region $z<0$ filled with a perfect conductor. We start by considering the field energy density in the vacuum half-space $z>0$, in particular in the very proximity of the vacuum-metal interface. The electric and magnetic field fluctuations, proportional to the corresponding energy densities, are well known in the literature for $z \neq 0$ (see \cite{SF02}, for example)

\begin{eqnarray}
\langle E^2(z) \rangle_R &=& \frac{3c\hbar}{4\pi z^4} \, ,
\label{eq:1}
\\
\langle B^2(z) \rangle_R &=& - \frac{3c\hbar}{4\pi z^4} \, ,
\label{eq:2}
\end{eqnarray}
where the $R$ subscript indicates that the space-independent energy density, present even in the absence of the metallic wall, has been subtracted (we will call them \emph{renormalized} field fluctuations).

Both electric and magnetic energy densities \eqref{eq:1} and \eqref{eq:2} decrease as $z^{-4}$ with the distance $z$ from the surface and diverge at the surface. Both the total (space integrated) electric and magnetic field energies diverge if considered independently, due to the behaviour of their densities at $z=0$.
Such a divergence is somehow related to the idealized assumptions of perfect conductor and of sharp boundary. It has been shown that even considering a real conductor described by the plasma model does not eliminate the divergence at the surface, because the plasma model does not give a sufficiently rapid convergence of the frequency integrals \cite{SF02}.

We will now show that a more detailed investigation of the energy densities for $z \to 0$ allows to point out the presence of extra singular terms at the interface position $z=0$, giving an essential contribution to the space-integrated electric and magnetic energies of the field, and thus to the wall self-energy \cite{BP12}. In other words, we show that the expressions \eqref{eq:1} and \eqref{eq:2} are not valid for $z=0$ and that extra terms exist.

In order to obtain the electric and magnetic energy densities, we start by considering, in place of the metallic one, a nondispersive and nondissipative dielectric half-space, with the dielectric characterized by a real and frequency-independent refractive index $n$; then, we take an appropriate limit to recover the ideal conductor case. It is this limiting procedure and the introduction of a cutoff that allow us to obtain the extra singular term at the boundary.
The vacuum-dielectric interface is the $xy$ plane and $z>0$ characterizes the vacuum region. The quantization of the electromagnetic field for such a system can be performed in terms of the Carniglia-Mandel triplets \cite{CM71}. Starting from a finite value of $n$, even if we then consider the limit $n \to \infty$, is mathematically useful in performing the calculation of renormalized quantities with a finite cutoff, allowing us to use the Carniglia-Mandel triplets as complete set of field modes \cite{BP12}. Details of the explicit evaluation of the vacuum electric and magnetic energy densities are given in \cite{HL99,BP12}, using a time splitting procedure, by introducing the following quantities
\begin{eqnarray}
\langle \mathbf{E}^2 (\br )\rangle_\eta= \sum_\lambda \langle \mathrm{E}_\lambda(\mathbf{r},t)\,\mathrm{E}_\lambda(\mathbf{r},t'=t+i\eta)\rangle \, ,
\label{eq:3} \\
\langle \mathbf{B}^2 (\br )\rangle_\eta= \sum_\lambda \langle \mathrm{B}_\lambda(\mathbf{r},t)\,\mathrm{B}_\lambda(\mathbf{r},t'=t+i\eta)\rangle \, ,
\label{eq:4}
\end{eqnarray}
with $\eta >0$ and the subscript $\lambda$ indicating the field components. This is exactly equivalent to introducing an exponential ultraviolet cutoff in the frequency integrals. In fact, when the expressions of the energy densities given above are explicitly written as a frequency integral, a function $e^{-\omega \eta}$ appears inside the integrals due to the time splitting, giving an ultraviolet cutoff for $\omega > 1/\eta$. Since we are only interested to these quantities in the ideal conductor limit $n \to \infty$, we report here only the expressions in this limit and in the vacuum limit $n \to 1$, necessary to obtain regularized quantities.

After lengthy algebraic calculations, we obtain for the electric field fluctuations
\begin{equation}\label{eq:5}
\langle\mathbf{E}^2\rangle_\eta^{\mathrm{vac}}=
\lim_{n\to 1}\langle\mathbf{E}^2\rangle_\eta=
\frac{12 \hbar }{\pi c^3 \eta ^4}
\end{equation}
for the vacuum case, and
\begin{equation}\label{eq:6}
\langle\mathbf{E}^2\rangle_\eta^{\mathrm{con}}=
\lim_{n\to \infty}\langle\mathbf{E}^2\rangle_\eta=
\frac{12 \hbar }{\pi c^3 \eta ^4}+\frac{4c\hbar}{\pi} \frac{\left(12 z^2-c^2 \eta^2\right)}{\left(4 z^2+c^2 \eta ^2\right)^3}
\end{equation}
for the ideal conductor case.

We can also obtain analogous expressions for the magnetic field fluctuations
\begin{equation}\label{eq:7}
\langle\mathbf{B}^2\rangle_\eta^{\mathrm{vac}}
=\lim_{n\to 1}\langle\mathbf{B}^2\rangle_\eta
= \frac{12 \hbar }{\pi c^3 \eta ^4} \, ,
\end{equation}
\begin{equation} \label{eq:8}
\langle\mathbf{B}^2\rangle_\eta^{\mathrm{con}}
=\lim_{n\to \infty}\langle\mathbf{B}^2\rangle_\eta
= \frac{12 \hbar }{\pi c^3 \eta ^4}-\frac{4c\hbar}{\pi} \frac{\left(12 z^2- c^2 \eta^2\right)}{\left(4 z^2+c^2 \eta ^2\right)^3}.
\end{equation}

Renormalized expressions are calculated by subtraction of the space-independent vacuum contributions \eqref{eq:5} and \eqref{eq:7} from \eqref{eq:6} and \eqref{eq:8}, respectively
\begin{equation}\label{eq:9}
\langle\mathbf{E}^2\rangle_{\eta \mathrm{R}}^{\mathrm{con}}=
\lim_{n\to \infty}\langle\mathbf{E}^2\rangle_\eta - \lim_{n\to 1}\langle\mathbf{E}^2\rangle_\eta
=\frac{4c\hbar}{\pi} \frac{\left(12 z^2- c^2 \eta^2\right)}{\left(4 z^2+c^2 \eta ^2\right)^3},
\end{equation}
\begin{equation}\label{eq:10}
\langle\mathbf{B}^2\rangle_{\eta \mathrm{R}}^{\mathrm{con}}=
\lim_{n\to \infty}\langle\mathbf{B}^2\rangle_\eta - \lim_{n\to 1}\langle\mathbf{B}^2\rangle_\eta
=-\frac{4c\hbar}{\pi} \frac{\left(12 z^2- c^2 \eta^2\right)}{\left(4 z^2+c^2 \eta ^2\right)^3}.
\end{equation}

Expressions \eqref{eq:9} and \eqref{eq:10} are finite for any nonvanishing value of $\eta$. A divergence at the conductor-vacuum interface $z=0$ appears in the limit $\eta \to 0$, that is for the cufoff frequency $1/\eta$ to infinity; in this limit, \eqref{eq:9} and \eqref{eq:10} give back, respectively, the $z^{-4}$ (with $z \neq 0$) ideal-conductor expressions \eqref{eq:1} and \eqref{eq:2}. The specific dependence of the energy densities around $z \sim c\eta$ depends in general on the specific cutoff function, and thus it is dependent on the model used for the metal. This is however not relevant for our purposes, because we are essentially interested in the ideal conductor limit $1/\eta \to \infty$, which allows us to make evident the existence of extra singular terms around $z=0$.

\begin{figure}[h]\centering
\includegraphics[width=8.7cm]
{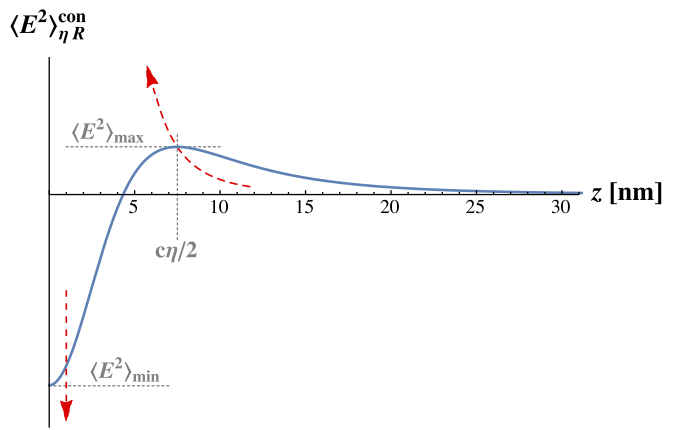}
\caption{(Color online) Renormalized electric field fluctuations $\langle\mathbf{E}^2\rangle_{\eta \mathrm{R}}^{\mathrm{con}}$ near the conductor-vacuum boundary as a function of the distance $z$ from the interface, for the value $1/\eta = 2 \cdot 10^{16} \, \mathrm{s}^{-1}$ of the cutoff frequency. The figure shows the maximum
$\langle E^2 \rangle_\text{max}=\frac{c\hbar}{\pi}\frac{1}{\eta^4c^4}$ at $z=c\eta /2$ and the minimum
$\langle E^2 \rangle_\text{min}=-\frac{c\hbar}{\pi}\frac{4}{\eta^4c^4}$ at $z=0$. The red arrows indicate how the maximum and minimum move when the cutoff frequency $1/\eta$ increases.}
\label{Fig:1}
\end{figure}

Figure \ref{Fig:1} shows $\langle\mathbf{E}^2\rangle_{\eta \mathrm{R}}^{\mathrm{con}}$, which is proportional to the renormalized electric energy density, near the interface as a function of $z>0$, for the cutoff frequency $1/\eta = 2 \cdot 10^{16} \, \mathrm{s}^{-1}$ of a typical metal. It shows that the electric energy density is finite at the interface, has a maximum close to the interface, and it decreases afterwards as $z^{-4}$ for larger distances.

It is straightforward to see that the integrals over $z$ of the renormalized energy density obtained from
\eqref{eq:9} and \eqref{eq:10} vanish for any nonvanishing value of $\eta$, even in the limit $\eta \to 0$. This behavior is completely different with respect to the space integrals of \eqref{eq:1} and \eqref{eq:2}, which diverge; this result stresses the importance of the extra terms we have obtained in the evaluation of the total electric and magnetic energies of the field (electromagnetic self-energy) from the corresponding densities.

It is worth to consider how the energy density behaves at the interface when the cutoff frequency $\eta^{-1}$ is increased and its limit to infinity (ideal conductor) is taken. We focus on the electric component; the same arguments hold for the magnetic one too.
$\langle\mathbf{E}^2\rangle_{\eta\ \mathrm{R}}^{\mathrm{con}}$ has a maximum at $z_\eta^{\mathrm{max}}=\frac{\eta c}{2}$ with the positive value $\frac{c\hbar}{\pi}\frac{1}{\eta^4c^4}$, and a minimum at $z_\eta^{\mathrm{min}}=0$ with the negative value  $-\frac{c\hbar}{\pi}\frac{4}{\eta^4c^4}$ (see Fig. \ref{Fig:1}). The width of the curve is about $0.5\eta c$.
For an increasing cutoff frequency $\eta^{-1}$, the curve in Figure \ref{Fig:1} approaches the $z^{-4}$ ideal conductor limit for distances from the interface larger than $c\eta$, but significant differences still remain close to the surface. In fact, while the ideal conductor result $\langle\mathbf{E}^2\rangle_\mathrm{R}^{\mathrm{con}}$ diverges with positive values at the surface (see Eq. \eqref{eq:1}), $\langle\mathbf{E}^2\rangle_{\eta\ \mathrm{R}}^{\mathrm{con}}$ assumes more and more negative values as $\eta^{-1} \to \infty$, and the width of the curve reduces to zero. Both maximum and minimum of $\langle\mathbf{E}^2\rangle_{\eta\ \mathrm{R}}^{\mathrm{con}}$ tend to collapse at the surface in the ideal conductor limit, yielding a surface divergence containing a nonvanishing electric and magnetic energy. Moreover, the renormalized electric fluctuation is negative at the exact position of the boundary, contrarily to the positive value predicted by \eqref{eq:1}.

The remarkable features of the electric (and magnetic) energy densities, that we have pointed out by exploiting our dielectric to ideal conductor limiting procedure, are not obtained when they are directly evaluated in the ideal conductor case (in this case only the terms in \eqref{eq:1} and \eqref{eq:2} are obtained).
In particular, as mentioned above, the negative minimum at the interface $z=0$ shown in Fig. \ref{Fig:1}, yielding a strong suppression of the electric energy density and field fluctuations, is a remarkable feature of the renormalized electric energy density we have obtained (this feature is not obtained if it is directly calculated for the ideal conductor as in \eqref{eq:1}). This may be relevant also because possibility and importance of suppression of vacuum energies, that is obtaining negative values of the renormalized energy density, has been recently raised in the literature \cite{Ford10}.
The result we have obtained shows that an intriguing structure of the electric and magnetic energy densities is present at the metal-vacuum interface.
They also demonstrate a discontinuity of the field energy densities at the boundary in the $\eta \to 0$ limit.

Finally, we wish to point out that the electric and magnetic energy densities and field fluctuations near the interface can in principle be experimentally investigated through the retarded Casimir-Polder interaction with an appropriate electrically and magnetically polarizable body \cite{PP87}. However, for a realistic absorbing medium, there will be other competing near-field contributions to the field energy density related to the imaginary part of the electric permittivity, which can also give important contributions in the very proximity of the interface.

\section{\label{sec:3}Singular behaviour of the field energy density near a point-like field source}

In the case discussed in the previous Section, the presence of the dielectric/metallic boundary was introduced through the boundary conditions it imposes on the field operators. We now consider the case of a point-like field source interacting with the quantum electromagnetic field in the ground state of the coupled system. In this case the source actively participates to the dynamics of the system. Also in this case, divergences of electric and magnetic energy densities are present at the position of the point-like source and they contain a finite amount of energy which contributes to the electromagnetic self-energy of the source \cite{PRS13}. Moreover, we discuss how the finite field energy arising from these singular terms is essential to ensure that the space integral of the field energy density correctly coincides with the total field energy obtained as the average value of the field Hamiltonian.

We suppose the source (a polarizable body or an atom, for example) located at $\br =0$ and describe its interaction with the field by
a simple effective Hamiltonian model, frequently used in the calculation of retarded interatomic Casimir-Polder forces \cite{CP69,PPT98}
\begin{eqnarray}
\label{eq:11}
H = \sum_{\bk j} \hbar \wk a^{\dagger}_{\bk j}a_{\bk j} + H_S -\frac 1 2 \alpha \mathbf{E}^2(0) \, ,
\end{eqnarray}
where $H_S$ is the Hamiltonian of the source and $\mathbf{E}$ is the electric field operator. We indicated by $\alpha$ the electric polarizability of the isotropic source that we assume real (nondissipative) and independent from the frequency (nondispersive). In this coupling scheme, the transverse displacement field, outside the source, coincides with the total (transverse plus longitudinal) electric field. Also, in this model for the source-field coupling, the source's internal degrees of freedom are \emph{frozen}, similarly to the macroscopic case of a conducting or dielectric boundary, or the case of a static source in quantum field theory \cite{HT62}.

The dressed ground state of the system, at the first order in the source's polarizability, is given by
\begin{eqnarray}
 \label{eq:12}
 \kvac&=&\mid g, 0_{\bk j} \rangle -\frac{\pi \alpha}{V}\sum_{\bk j}\sum_{\bkp j'} \frac{(kk')^{1/2}}{k+k'} \hat{e}_{\bk j}\cdot\hat{e}_{\bkp j'} \nonumber \\
 &\times& \mid g, 1_{\bk j}1_{\bkp j'}\rangle ,
\end{eqnarray}
where $g$ denotes the source ground state, $\hat{e}_{\bk j}$ are polarization unit vectors and $V$ is the quantization volume. The true ground state \eqref{eq:12} contains mixtures with states containing virtual photons, which modify the field fluctuations and energy densities in the space around the source.

After some algebra, the renormalized electric and magnetic energy densities around the source are obtained as
\begin{eqnarray}
\label{eq:13}
& & \frac{1}{8\pi}  \bvac \mathbf{E}^2(\br) \kvac \nonumber \\
&=&\frac{\alpha\hbar c}{4\pi^3}\int_{0}^{\infty} \! \! dk \int_0^{\infty} \! \! dk'
\left[ j_0(kr)j_0(k'r)- j_0(kr)\frac{j_1(k'r)}{k'r}\right. \nonumber \\
&-& \left.  \frac{j_1(kr)}{kr}j_0(k'r)+\frac{3}{k k' r^2} j_1(kr)j_1(k'r) \right]
\frac{k^3 k'^3}{k+k'} \, , \\
\label{eq:14}
& &\frac{1}{8\pi}  \bvac\mathbf{B}^2 (\br) \kvac \nonumber\\
&=&-\frac{\alpha\hbar c}{2\pi^3}\int_{0}^{\infty}  \! \! dk \int_0^{\infty} \! \! dk'
 j_1(kr)j_1(k'r) \frac{k^3 k'^3}{k+k'} \,
\end{eqnarray}
where $j_n(x)$ are spherical Bessel functions. The $k,k'$ integrals in \eqref{eq:13} and \eqref{eq:14} can be factorized using the relation
$(k+k')^{-1}=\int_0^{\infty} \! \! e^{-(k+k')\gamma} d\gamma$, obtaining
\begin{eqnarray}
\label{eq:15}
\frac{1}{8\pi}  \bvac \mathbf{E}^2(\br) \kvac &=& \frac{4\alpha\hbar c}{\pi^3}
\int_0^{\infty} \! \! d\gamma\frac{3 r^4-2r^2\gamma^2+3\gamma^4}{(r^2+\gamma^2)^6},\\
\label{eq:16}
\frac{1}{8\pi}  \bvac\mathbf{B}^2 (\br) \kvac &=&-\frac{4\alpha\hbar c}{\pi^3}
\int_0^{\infty}  \! \! d\gamma\frac{8 r^2\gamma^2}{(r^2+\gamma^2)^6} \, .
\end{eqnarray}

The integration over $\gamma$ in \eqref{eq:15} and \eqref{eq:16} must be performed carefully in order to deal properly with their behaviour at the source's position $r=0$.  For $r \neq 0$, Eqs. \eqref{eq:15} and \eqref{eq:16} immediately yield the $r^{-7}$ behaviour of both the electric and magnetic energy densities \cite{CPP95}.

A more careful procedure to calculate \eqref{eq:15} and \eqref{eq:16}, valid also at the source position, involves the introduction of an exponential cut-off function in the integrals, and taking the limit of the cut-off frequency to infinity only after performing the integrals over $\gamma$ \cite{PRS13}. In this case, we have
\begin{eqnarray}
\label{eq:17}
& &\frac{1}{8\pi}  \bvac \mathbf{E}^2(\br) \kvac  = \frac{\hbar c \alpha}{(4\pi)^2}\left\{ \frac{23}{r^7}-\frac{23}{r^6} \delta (r) +\frac{10}{r^5}\delta'(r)\right.\nonumber\\
& &\left.-\frac{7}{3r^4}\delta''(r)
+\frac{1}{3r^3}\delta'''(r)+\frac{1}{15r^2}\delta^{(iv)}(r)\right\} , \\
\label{eq:18}
& &\frac{1}{8\pi}  \bvac\mathbf{B}^2 (\br) \kvac = -\frac{\hbar c \alpha}{(4\pi)^2}\left\{ \frac{7}{r^7}-\frac{7}{r^6} \delta (r) +\frac{2}{r^5}\delta'(r)\right.\nonumber\\
& &\left.+\frac{1}{3r^4}\delta''(r)
-\frac{1}{3r^3}\delta'''(r)-\frac{1}{15r^2}\delta^{(iv)}(r)\right\} ,
\end{eqnarray}
where the superscript to the delta function indicates the order of its derivative with respect to $r$. The first term on the right-hand side of \eqref{eq:17} and \eqref{eq:18} is the well-known $r^{-7}$ contribution yielding the far-zone electric-electric and electric-magnetic interatomic Casimir-Polder potential energy \cite{PP87,CPP95}. The other terms are new singular terms localized at the position of the source; they give a finite contribution to the space integral of the electric and magnetic energy densities and thus to the electromagnetic self-energy of the source.

It is possible to show from Eqs. \eqref{eq:15} and \eqref{eq:16}, or from \eqref{eq:17} and \eqref{eq:18}, that the integral over all space of the total energy density of the field
$\bvac\mathcal{H}_{el}(\br) \kvac + \bvac \mathcal{H}_{m}(\br) \kvac$ vanishes
\begin{equation}
\label{eq:18a}
\int \! d^3 r \bvac \left[ \mathcal{H}_{el}(\br) + \mathcal{H}_{m}(\br) \right] \kvac = 0 \, .
\end{equation}
We wish to point out that the existence of the integral \eqref{eq:18a} confirms that the field energy density as obtained from \eqref{eq:17} and \eqref{eq:18} is a mathematically well-defined quantity as a distribution.

The fact that the integral over all space of the total energy density (electric plus magnetic) vanishes, as shown by Eq. \eqref{eq:18a}, is consistent with the vanishing value of the field energy as obtained by taking the average value of the field Hamiltonian in \eqref{eq:11} on the dressed ground state \eqref{eq:12}.
If the singular terms in $r=0$ in \eqref{eq:17} and \eqref{eq:18} containing the delta function and its derivatives were neglected (that is, considering only the usual $r^{-7}$ terms), the space integral of the field energy would have been diverging due to its behavior at $r=0$; in such a case, a sharp inconsistency between local and global self-energy would have appeared. This is a further aspect showing the importance of properly dealing with the singular behavior of the field energy densities at a boundary or at the position of field sources, in order to gain a mathematical consistency of the theory.

It has been also pointed out that the singularities we have discussed are smeared out if extended field sources are considered \cite{PRS13}, even if we can expect that a significant concentration of the energy density remain in the proximity of the extended source. We expect that such a smearing out of the singularity should occur also if we consider field sources with a fluctuating position, similarly to what happens in the case of a mobile boundary considered in the next Section.

\section{\label{sec:4}Field energy density near a mobile wall and surface divergences}

In the previous Sections we discussed the presence of divergences of the field energy density near a vacuum-metal interface or a point-like field source, and their importance from a physical point of view. In this Section we will show that such a kind of singular behavior is smoothened once such a boundary is allowed to move in space.
Systems where field modes are coupled to cavities with moving walls are typically considered in the growing field of quantum optomechanics, dealing with the coupling between field (optical) and mechanical degrees of freedom \cite{AKM13,Meystre13,Chen13}.
The relevance of systems analogous to the one we deal with in this Section is also related to the very intriguing possibility, recently suggested in the literature, for the motion of the mirror, or other internal degrees of freedom, to provide the actual value of the vacuum energy density independently of possible gravitational effects \cite{WU14}.

We consider the simple case of a 1D massless scalar field $\phi (x,t)$ between two perfect mirrors placed at $x=0$ and $x=L(t)$. The first mirror is fixed in space, while the second one can move and is bounded to its equilibrium position $L_0$ by a harmonic potential $V(q)$. The mobile mirror's degrees of freedom are treated quantum mechanically.
The motion of the mirror couples its mechanical degrees of freedom to the field modes inside the cavity, and drives an effective interaction between the field modes. The mirror is thus subject to quantum fluctuations of its position, due to both uncertainty principle of its mechanical degrees of freedom and radiation pressure. The moving wall imposes time-dependent Dirichlet boundary conditions to the field operator, $\phi (0,t)=\phi(q(t),t)=0$ (perfect mirror), and is in turn subjected to the relative radiation pressure: these conditions give origin to the coupling between mechanical and field degrees of freedom. Canonical quantization of this system has been obtained in Refs. \cite{Law94,Law95} in terms of an effective Hamiltonian, assuming a small displacement of the moving mirror from its equilibrium position. Such effective Hamiltonian has the form
\begin{eqnarray}
 H &=& \hbar \wo b^\dag b +\hbar \sum_k \omega_k a_k^\dag  a_k \nonumber \\
 &-& \sum_{kj}C_{kj} \Big\{ \left( b + b^\dag \right)
{\cal N} \! \left[ \left( a_k + a_k^\dag \right) \left(  a_j + a_j^\dag \right) \right] \Big\} \, ,
\label{eq:19}
\end{eqnarray}
where ${\cal N}$ is the normal ordering operator and
\begin{equation}
C_{kj}=(-1)^{k+j}\left(\frac \hbar 2\right)^{3/2}\frac 1{L_0 \sqrt{M}}\sqrt{\frac{\omega_k \omega_j}\wo}
\label{eq:20}
\end{equation}
is the coupling constant between field and mobile mirror. We have indicated by $M$ the mass of the mobile mirror, while $a_k, \, a_k^\dagger$ are annihilation and creation operators of the field modes between the two mirrors relative to the equilibrium position $L_0$ with wavenumber $k$, and $b, \, b^\dagger$ are annihilation and creation operators relative to the mobile wall (they describe excitations of the wall's mechanical degrees of freedom). In our 1D model, the field modes are equally spaced in frequency; the allowed frequencies are $\omega_j = c\kappa_j$, with $\kappa_j=(\pi /L_0)j$, $j$ being an integer number.

In Hamiltonian \eqref{eq:19}, field quantization has been performed in terms of modes relative to the wall's equilibrium position $L_0$. In \cite{BP13} we have used this Hamiltonian to describe the zero-point field fluctuations inside the cavity and their modification due to the wall's motion (position fluctuations), as well the consequent change of the Casimir force between the two mirror and of the Casimir-Polder force on a polarizable body.

Due to the effective interaction between the mobile wall and the field, described by the Hamiltonian \eqref{eq:19}, the true ground state of the interacting system is dressed by virtual excitations both in the field and in the wall. At the lowest significant order in the interaction, the dressed ground state is given by \cite{BP13}
\begin{equation}
\ket{g}=\ket{\left\{0_k\right\},0}+\sum_{kj}{D_{kj}\ket{\left\{1_k,1_j\right\},1}} \, ,
\label{eq:21}
\end{equation}
where
\begin{equation}
D_{kj}=(-1)^{k+j}\frac 1{L_0} \sqrt{\frac {\hbar \wk \wj}{8 M \wo}}\frac 1{\wo+\wk+\wj}.
\label{eq:22}
\end{equation}
In Eq. \eqref{eq:21}, the first element of the state vectors (inside the curly bracket) refers to field excitations, while the second element refers to excitations of the wall's mechanical degrees of freedom.

The state \eqref{eq:21} contains terms with two excitations in the field and one excitation in the wall, analogously to the dynamical Casimir effect where pairs of real photons are emitted by the oscillating wall \cite{Dodonov10}; in the present case, however, the field excitations are virtual quanta and,  furthermore, the motion of the wall follows from the internal dynamics of the system described by the Hamiltonian \eqref{eq:19}, and it is not given by an external action as in the usual dynamical Casimir effect. The presence of virtual quanta of the field in the dressed ground state yields a change of the zero-point fluctuations and field energy density in the cavity with respect to the case of fixed walls \cite{BP13}, analogously to the case of the field source considered in the previous Section.

We can evaluate the field energy density inside the cavity, by evaluating the average value of the field energy density operator
\begin{equation}
\mathcal{H}(x)=\frac 12 \left[\frac 1{c^2} {\dot{\phi}}^2(x)+\left( \frac {d {\phi (x)}}{dx} \right)^2 \right]
\label{eq:23}
\end{equation}
on the dressed ground state \eqref{eq:21}. The \emph{renormalized} energy density, obtained after subtraction of the space-independent contribution present even in the absence of the cavity walls
\begin{equation}
\langle \left\{0_k\right\} \mid \mathcal{H} \mid \left\{0_k\right\} \rangle = -\frac {\pi  c \hbar}{24 L_0^2} \, ,
\label{eq:23a}
\end{equation}
is given by
\begin{eqnarray}
&\ & \langle \mathcal{H}^{\text{R}}(x) \rangle =  \langle g \mid \mathcal{H}^{\text{R}}(x) \mid g \rangle \nonumber \\
&\ & =\langle g \mid \mathcal{H} \mid g \rangle
- \langle \left\{0_k\right\} \mid \mathcal{H} \mid \left\{0_k\right\} \rangle
\nonumber \\
&\ &= \frac{\hbar^2}{2L_0^3 M \wo} \sum_j \Big\{ \sum_{kp} (-1)^{k+p} \cos [(\kappa_k-\kappa_p)x]
\nonumber \\
&\ & \times \frac {\omega_k \omega_j\omega_p}{(\wo+\omega_k+\omega_j)(\wo +\omega_p+\omega_j)} \Big\} \, ,
\label{eq:24}
\end{eqnarray}
where $\kappa_k = k\pi /L_0$, with $k$ an integer number.

The quantity \eqref{eq:24} can be evaluated numerically. The sum over $j$ has an ultraviolet divergence, that can be cured by introducing an upper cutoff frequency $\wc$ to regularize the sums over frequencies. A typical value of this cutoff frequency can be the plasma frequency of the mirror. Eq. \eqref{eq:24} shows that the change of the energy density significantly depends on the oscillation frequency of the mirror $\wo$ and its mass $M$, scaling as $1/M$.

\begin{figure}[h]\centering
\includegraphics[width=8cm]
{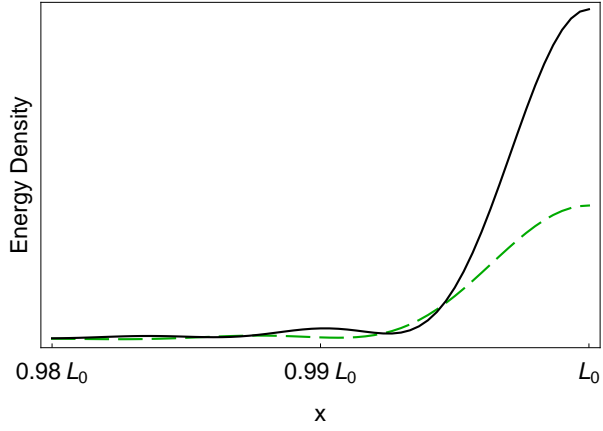}
\caption{(Color online) Change of the renormalized field energy density (arbitrary units), compared to the static walls case, in the proximity of the moving mirror, where the effect is particularly relevant. The two curves differ for the value of the cutoff frequency: $\wc = 8 \cdot 10^{15} \, \text{s}^{-1}$ (green dashed line) and $\wc = 10^{16} \, \text{s}^{-1}$ (black continuous line). The other parameters used are:
$\wo = 10^5 \, \text{s}^{-1}$, $L_0 = 10 \, \mu \text{m}$ and $M=10^{-11}\, \text{kg}$.}
\label{Fig:2}
\end{figure}

Fig. \ref{Fig:2} shows the dependence of the renormalized field energy-density change \eqref{eq:24} on the position inside the cavity in the very proximity of the mobile wall, where the effect is more relevant. The result for two different values of the cutoff frequency, $8 \cdot 10^{15 }\, \text{s}^{-1}$ and $10^{16 }\, \text{s}^{-1}$, are compared; the other parameters used are: $\wo = 10^5 \, \text{s}^{-1}$, $L_0 = 10 \, \mu \text{m}$ and $M=10^{-11}\, \text{kg}$. The figure clearly shows that the change of the field energy density is relevant in the proximity of the mobile wall (average position $x=L_0$), while it becomes negligible far from it. Also, in correspondence of the mobile wall's equilibrium position $L_0$, it increases with the cutoff frequency, and it is possible to show that it diverges at $x=L_0$ for $\wc \to \infty$ (perfect mirror), similarly to the case considered in Section \ref{sec:2}. From a physical point of view, the fact that the change of the energy density is concentrated near the mobile wall can be easily understood by taking into account that the virtual quanta in \eqref{eq:21} cannot propagate and remain confined near the wall according to the time-energy uncertainty relation. This increase of the energy density, due to the movement of the wall (position fluctuations) can have important observable consequences because of possible significant gravitational effects, the field energy density acting as a source term for the gravitational field.

The effect we have considered grows with a decreasing mass of the mobile wall as it is evident from \eqref{eq:24}; masses so small as $10^{-21} \, \text{kg}$ are currently reachable by modern optomechanical techniques \cite{AKM13}, making possible observation of this phenomenon with actual techniques \cite{BP13}. It should be also noted that vacuum energy density and local field fluctuations can be experimentally probed through the Casimir-Polder interaction with a polarizable body. In analogy with the case of electric field fluctuations, we can write the interaction energy with a polarizable body as $-1/2 \alpha {\dot{\phi}}^2(x)$, where $\alpha$ is the static polarizability; the analogy with the magnetic field fluctuations gives an analogous expression involving $(d \phi (x)/dx)^2$ and an equivalent of the magnetic polarizability) \cite{PP87}.

As mentioned in the previous Sections, it has been suggested in the literature that the problem of surface divergences at the interface could be solved by imperfect or fluctuating boundaries \cite{FS98}; although in our case the boundary has indeed quantum position fluctuations, the singular behavior of the energy density for $\wc \to \infty$ is still present because in the effective Hamiltonian \eqref{eq:19}, the field is quantized in terms of operators relative to the wall's equilibrium position.

We now show that this singular behaviour of the field energy density is naturally softened and smeared out when the position fluctuations of the mobile wall, which are included in our dressed ground state \eqref{eq:21}, are properly taken into account by a statistical averaging.

From \eqref{eq:21} we can obtain the wall's reduced density operator
\begin{equation}
\rho_\text{osc}= \left( 1-N_b \right) \mid 0 \rangle \langle 0 \mid + N_b \mid 1 \rangle \langle 1 \mid \, ,
\label{eq:25}
\end{equation}
where
\begin{equation}
N_b = \frac \hbar{4L_0^2M} \sum_{kj}\frac {\wk \wj}{\wo (\wo +\wk +\wj)^2}
\label{eq:26}
\end{equation}
is the average number of excitation of the mobile wall in the ground state. The position distribution $f(q)$ of the wall is thus given by a statistical mixture of probabilities relative to a ground-state and first-excited-state of an harmonic oscillator
\begin{equation}
f(q) = (1-N_b) f_0(q) + N_b f_1(q) \, ,
\label{eq:27}
\end{equation}
with
\begin{eqnarray}
f_0(q) &=& \left( \frac {M\wo}{\pi\hbar}\right)^{1/2}e^{-\frac{M\wo}h q^2} \, , \\
f_1(q) &=& \left[ \frac 4\pi \left( \frac {M\wo}\hbar\right)^3\right]^{1/2}q^2 e^{-\frac{M\wo}h q^2} \, .
\label{eq:28}
\end{eqnarray}

The position distribution probability \eqref{eq:27} follows from the internal dynamics of the system, and it is not externally prescribed.
We can now average the energy density \eqref{eq:24} over the distribution probability of the mirror's position given by \eqref{eq:27}, assuming small position fluctuations compared to $L_0$. After some straightforward algebra, we finally obtain
\begin{eqnarray}
\langle \langle \mathcal{H}^{\text{R}}(x) \rangle \rangle &=&  \frac {\hbar^2}{2L_0^3 M \wo} \sum_{jkp} (-1)^{k+p}
\nonumber \\
&\times& \! \! \frac {\omega_k \omega_j\omega_p}{(\wo+\omega_k+\omega_j)(\wo +\omega_p+\omega_j)}
\nonumber \\
&\times& \! \!  \sum_{n=0}^1p_n \langle \cos [( \kappa_k -\kappa_p )x] \rangle_n \, ,
\label{eq:29}
\end{eqnarray}
where $p_0=1-N_b$, $p_1=N_b$, and
\begin{eqnarray}
&\ & \langle \cos [( \kappa_k -\kappa_p )x] \rangle_0 = \int_{-\infty}^\infty \!\! dq \cos [( \kappa_k(q) -\kappa_p(q) )x] f_0(q)
\nonumber \\
&\simeq& e^{-\frac {\hbar (\kappa_k -\kappa_p )^2x^2}{4L_0^2M\wo}} \cos  [( \kappa_k -\kappa_p )x] \, ,
\label{eq:30}
\end{eqnarray}
\begin{eqnarray}
&\ &\langle \cos [( \kappa_k -\kappa_p )x] \rangle_1 = \int_{-\infty}^\infty \!\! dq \cos [( \kappa_k(q) -\kappa_p(q) )x] f_1(q)
\nonumber \\
&\simeq& \! \! e^{-\frac {\hbar (\kappa_k -\kappa_p )^2x^2}{4L_0^2M\wo}}
\left[ 1 - \frac {\hbar (\kappa_k -\kappa_p )^2x^2}{2L_0^2M\wo} \right]
\cos  [( \kappa_k -\kappa_p )x] \, .
\label{eq:31}
\end{eqnarray}

\begin{figure}[h]\centering
\includegraphics[width=8cm]
{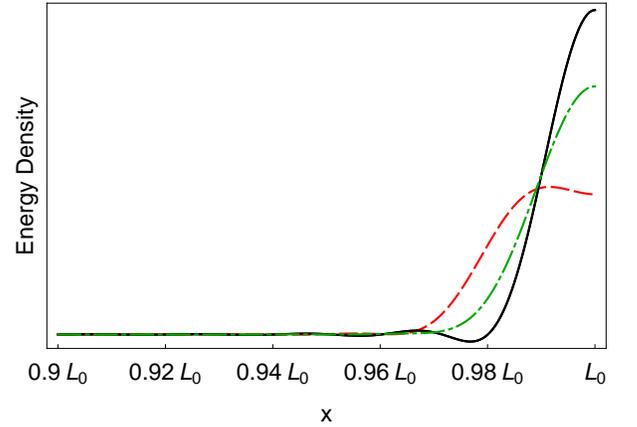}
\caption{(Color online) Comparison between the change of the renormalized field energy density (arbitrary units) in the proximity of the mobile wall in three different cases: non-averaged case (black continuous curve), average over the mirror's ground state (green dot-dashed curve) and average over the mirror's first excited state (red dashed curve). The number of modes is $100$, yielding a cutoff frequency $100c\pi /L_0$.}
\label{Fig3}
\end{figure}

\begin{figure}[h]\centering
\includegraphics[width=8cm]
{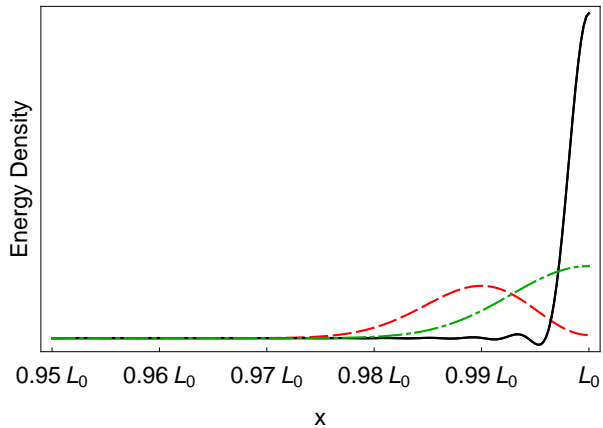}
\caption{(Color online) This curve shows the same quantities of Fig. \ref{Fig3} (arbitrary units) with 500 field modes, yielding a cutoff frequency equal to $500c\pi /L_0$.}
\label{Fig4}
\end{figure}

The quantity \eqref{eq:29} can be evaluated numerically using a finite set of cavity field modes. An upper cutoff frequency $\wc$ can be introduced on a physical basis because the walls become transparent for frequencies larger than their plasma frequency, as already done to evaluate the energy density \eqref{eq:24}. In our 1D model the modes are equally spaced in frequency by $c\pi /L_0$, and thus setting an upper cutoff is equivalent to set the number of modes. Figures \ref{Fig3} and \ref{Fig4} show the renormalized field energy density  \eqref{eq:29} for $100$ and $500$ modes, respectively, in the very proximity of the mobile wall. In this region, the leading contribution is mainly from the highest frequency modes.
Both figures show the energy density without averaging over the mirror's position (black continuous curve) and, for the averaged case, the two contributions in \eqref{eq:29} from $n=0$ and $n=1$  separately: the contribution coming from the average over the mirror's ground state (green dot-dashed curve), as obtained from the term with $n=0$ in \eqref{eq:29} and \eqref{eq:30}, and the contribution coming from the average over the mirror's first excited state (red dashed curve), coming from the term with $n=1$ in \eqref{eq:29} and \eqref{eq:31}. Parameters are chosen such that the square root of the average quadratic displacement of the wall is $L_0/100$. These results clearly show that the averaging process has softened the divergence of the field energy density at the mobile wall, smearing it over the space around the mirror's equilibrium position. Also, in the second case (500 modes), it is clear that the maximum value of the energy density it is not necessarily at the average position of the wall. This is consistent with the physical hypothesis that the singular behaviour at boundaries of the energy densities is eliminated in the realistic case of fluctuating boundaries.

\section{\label{sec:5}Conclusions}
We have discussed and reviewed some aspects of the singular behaviour of the vacuum field energy densities and field fluctuations in the proximity of a boundary or a point-like field source.

We have first considered the case of the interface between a perfectly conducting wall and the vacuum space. Using an appropriate limiting procedure from a dielectric to an ideal conductor, we have shown the existence of extra singular terms in this limit in both the electric and magnetic energy densities, apart the well-known $z^{-4}$ behavior, $z$ being the distance from the interface. These singular terms contain a finite amount of energy and thus significantly contribute to the self-energy of the system.
They could also yield relevant gravitational effects, because the energy density is a source term for gravity; the possible coupling of field fluctuations to gravity, however, is still an intriguing controversial subject in the literature. We have then considered a point-like source of the electromagnetic field, described in term of its polarizability and an effective Hamiltonian. We have shown that also in this case the electric and magnetic energy densities have singularities at the source's position, and that the singular terms we obtain have a finite energy. They contribute to the electromagnetic self-energy of the system, as in the previous case. Moreover, these singular terms are essential to ensure consistency between the evaluation of the renormalized field energy in the true ground state, as obtained from the space integral of its density or from the average of the field Hamiltonian in the dressed ground state.

It is however expected that these singular behaviours should be softened and smeared out for fluctuating boundaries and extended field sources. Thus, we have considered a massless scalar field in a 1D cavity with one fixed and one mobile wall, whose degrees of freedom are treated quantum-mechanically. The mobile wall has therefore quantum fluctuations of its position, due to both the uncertainty relation for its mechanical degrees of freedom and to vacuum radiation pressure. We have evaluated the field energy density inside the cavity in the proximity of the mobile wall, where the effects of its movement are expected to be larger. We have shown that, in the proximity of the mobile wall, the position fluctuations of the wall change the field energy density and that this change can be relevant for small mirror's masses and becomes singular with an increasing cutoff frequency. We have finally shown that a proper inclusion of the quantum-mechanical motion of the wall, by averaging the field energy density over the actual mirror's position distribution probability, smoothens and smears out the singularity of the energy density at the mobile boundary. This result confirms that including position fluctuations of a boundary can eliminate the singularity of the field energy density.

\begin{acknowledgments}
The authors gratefully acknowledge the Julian Schwinger Foundation for financial support. Financial support by MIUR, CRRNSM, SUPA and EPSRC CM-DTC Grant No EP/403673X/1 is also acknowledged.\end{acknowledgments}

\end{document}